\documentclass[aps,letterpaper,twocolumn]{revtex4-1}	
\usepackage{float}
\usepackage{graphicx,subfigure}
\usepackage{overpic}
\usepackage{cancel}
\usepackage{amsmath, amsfonts,amssymb}
\usepackage{bm}
\usepackage{sidecap}
\usepackage{color}
\usepackage{outlines}
\usepackage{soul}
\usepackage{hyperref}
\usepackage{dsfont}
\newcommand{\be}{\begin{equation}}
\newcommand{\ee}{\end{equation}}

\newcommand{\Tr}{{\rm Tr}}
\newcommand{\llangle}{\langle \!\langle}
\newcommand{\rrangle}{\rangle \! \rangle}

\newcommand{\Ket}[1]{|#1\rangle}
\newcommand{\Hes}{\hat{\boldsymbol{H}}}

\definecolor{violet}{rgb}{0.58, 0.0, 0.83}
\definecolor{lightblue}{rgb}{0.6, 0.6, 1}
\newcommand{\mkadd}[1]{{\color{violet}{#1}}}

\newcommand{\mkdel}[1]{}
\newcommand{\mkdelmath}[1]{}

\newcommand{\fnadd}[1]{{\color{blue}{#1}}}
\newcommand{\FNcomment}[1]{{\fnadd{\bf [fn: #1]}}}

\newcommand{\fnrm}[1]{\textcolor{blue}{{ }\st{#1}{ }}}

\newcommand{\fnmv}[1]{{\color{lightblue} [#1]}}

\renewcommand{\fnrm}[1]{}
\renewcommand{\fnmv}[1]{}
\renewcommand{\fnadd}[1]{\textcolor{blue}{#1}}

\renewcommand{\FNcomment}[1]{}

\newcommand{\zs}{\hspace{1.5pt}}
\newcommand{\bs}{\boldsymbol}
\begin{document}
\title{Quasiperiodic Floquet-Thouless energy pump}
\author{Frederik Nathan$^{1,2}$, Rongchun Ge$^3$, Snir Gazit$^4$, Mark Rudner$^1$, Michael Kolodrubetz$^3$}
\affiliation{$^1$Center for Quantum Devices, Niels Bohr Institute, University of Copenhagen, 2100 Copenhagen, Denmark}
\affiliation{$^2$Department of Physics and Institute for Quantum Information and Matter,
California Institute of Technology, Pasadena, CA 91125, USA}
\affiliation{$^3$Department of Physics, The University of Texas at Dallas, Richardson, Texas 75080, USA}
\affiliation{$^4$Racah Institute of Physics and The Fritz Haber Research Center for Molecular Dynamics, The Hebrew University of Jerusalem, Jerusalem 91904, Israel}

\begin{abstract}
We study a disordered one-dimensional fermionic system subject to quasiperiodic driving by two modes with incommensurate frequencies.
We show that the system supports a topological phase  in which energy is transferred between the two driving modes at a quantized rate. 
The phase  is  protected  by a combination of disorder-induced spatial localization and frequency localization, a mechanism unique to quasiperiodically driven systems.
We demonstrate that an 
analogue of the  phase can be realized in  a cavity-qubit system driven by two incommensurate modes.
\end{abstract}
\maketitle

 Periodic driving can be used as a tool for quantum control~\cite{QC,FQE1,FQE2,Lindner_2011_1, Struck12, Wang_2013_1, Jotzu_2014, McIver_2020,DDcou1,DDcou2,DDcou3,Fana0,Fana1,Fana2} and can even  induce new phases of matter  with no  equilibrium analogues~\cite{Jiang2011,FQI1,Nathan2015,PP1,PP2,PP3,FQI0,FQI1,FQI2,FQI3,FQI4,Wintersperger_2020,Sacha_2015,TC1,TC2,TC3,TC4}.
Recently, it was discovered that {\it quasiperiodically} driven systems  
 also support their own unique phases of matter~\cite{Else_2020}, despite having neither continuous nor discrete time-translation symmetry.

In this work, we report the discovery of  a  topological phase of matter 
in  quasiperiodically driven systems. 
We study a one-dimensional  (1d) fermionic system 
driven  by two modes with incommensurate frequencies, $\omega_1$ and $\omega_2$.
With  spatial disorder, 
the system supports a   phase which is characterized by quantized energy transport and 
a nonzero value of an integer-valued 
 topological invariant $\nu$. 
When  one end of the system is fully occupied by fermions  {while the other end is empty} [as  depicted in Fig.~\ref{f1}(a)], 
the system transfers energy  between the driving modes  at the quantized average rate $\nu \mathcal P_0$, where $\mathcal P_0 \equiv {\omega_1\omega_2}/{2\pi}$  denotes the ``quantum of energy transfer'' (with $\hbar=1$ throughout).
 We refer to this phase as the quasiperiodic Floquet-Thouless energy pump (QFTEP).

The absence of time-translation symmetry  gives the QFTEP   features which have no analogue in equilibrium or Floquet systems.
In particular, the QFTEP is protected by a combination of spatial and frequency 
localization~\cite{Ho_1983,Quasi_Luck88,Quasi_Casati89,Quasi_Blekher92,Hatami_2016,Else_2020}, meaning the index $\nu$ can only change if this localization is destroyed.
Here frequency localization 
 is a phenomenon unique to quasiperiodically driven systems which arises only for sufficiently irrational $\omega_2/\omega_1$. 
The condition of irrational $\omega_2/\omega_1$ means the QFTEP features  a fractally-structured phase diagram (see Fig.~\ref{f2} and discussion below for details).

The same ``quantum of energy transfer'' that we observe  was recently encountered in Refs.~\cite{TQEP1,TQEP2,Fexch}. 
In particular, Ref.~\cite{TQEP1} studied the same class of  systems we consider here,  in the limit 
 where the second driving mode is adiabatic: 
$\omega_2 \to 0$.
For fine-tuned parameters, and in the absence of disorder, this system was shown to exhibit a similar quantized energy pumping phenomenon to that we observe here. 
Unlike the phenomena in Refs.~\cite{TQEP1,TQEP2,Fexch},  the QFTEP does not require adiabatic driving,    is robust to   disorder,   and occupies a finite region of parameter space. 
Hence the QFTEP, in contrast to these earlier phenomena,  constitutes a genuine phase of matter.

\begin{figure}[!t]
\centering\includegraphics[clip=true,width=.99\columnwidth]{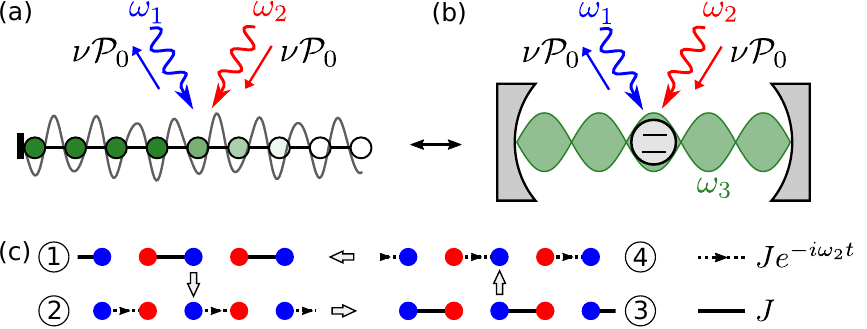}
\caption{(a) The quasiperiodic Floquet-Thouless energy pump for a fermionic chain driven by incommensurate frequencies $\omega_1$ and $\omega_2$.
 When one end of the chain is fully occupied (green), energy is pumped between the two modes at the rate $\nu \mathcal P_0$ ($\mathcal P_0 = \omega_1\omega_2/2\pi$), where $\nu \in \mathds{Z}$ is the topological winding number of the phase. The quantization persists as long as the system is localized in both real and frequency space.
 (b) {An analogue of} the  phase can be realized with a bichromatically driven qubit-cavity system. 
(c) 
$4$-step driving protocol that realizes the phase.
A particle tunneling along (against) 
the arrow acquires a phase $e^{-i\omega_2 t}$ ($e^{i\omega_2t}$).} 
\label{f1}
\end{figure}

We propose an experimental realization of a dimensionally-reduced
version of the QFTEP in a two-level system (qubit) coupled to a quantized  cavity mode and driven by two incommensurate frequencies [see Fig.~\ref{f1}(b)]. Our results indicate that such a simple physical setup inherits the topological properties of the QFTEP, 
suggesting the possibility of realizing this phase in cavity quantum electrodynamics.

{\it Model ---} 
{Here we present a particular 
 model that realizes the QFTEP. }
We consider a 1d bipartite tight-binding system with $L$ unit cells, 
with Hamiltonian 
$
\hat{\mathcal{H}}_0(t{; \lambda}) = \hat {\mathcal{H}}_{\rm dr}(t{; \lambda}) + \hat{\mathcal{H}}_\mathrm{dis}
\label{eq:model}
$. 
Here, for fixed $\lambda \in \mathbb{R}$, $\hat{\mathcal{H}}_{\rm dr}(t; \lambda)$ describes time-dependent 
 tunneling with period $T_1$, while $\hat{\mathcal{H}}_{\rm dis}= \sum_{x=1}^{2L}\frac{\Delta_{x }}{2} 
 \hat c^{\dagger}_{x} \hat c_{x}$
 describes a static on-site disorder potential.
 Here  $\hat c_{x}$ annihilates a fermion on site $x$,  
 while each $\Delta_{x }$ is picked randomly from the interval $[-W, W]$. We take the lattice constant to be $1$ throughout.
The driving term $\hat{\mathcal{H}}_{\rm dr}(t; \lambda)$  is piecewise constant in $t$ over $4$ steps of equal length [see Fig.~\ref{f1}(c)].
 In step $s$, defined as  the interval $\frac{s-1}{4}T_1\leq t < \frac{s}{4} T_1$,  $\hat{\mathcal{H}}_{\rm dr}(t; \lambda) = - J\sum_{n=1}^L (e^{i\eta _s\lambda}\hat c_{2n }^\dagger \hat c_{2n+d_s} + {\rm H.c.})$,  where   $d_{1}=d_4=  1$, $d_{2}=d_3= -1$, $\eta_4  =-\eta_2 = 1$, and $\eta_1=\eta_3 =0$. 
The parameter $\lambda$ controls the phases of the tunnelling terms. 
 We set $J=\frac{2\pi}{T_1}$ such that  $\hat{\mathcal{H}}_{\rm dr}(t; \lambda)$  generates tunneling by precisely one site per step when $\lambda$ is  fixed.

Related versions of the model above were  studied in  Refs.~\cite{FQI1,FQI2,TQEP1}.
Ref.~\onlinecite{TQEP1} explored the case where  $\lambda$ was  increased adiabatically, and   argued that this cyclic modulation 
    caused a transfer of energy to the driving mode at the quantized rate of $\omega_1 = 2 \pi /T_1$ per cycle.
In this work, we consider the case where $\lambda$ increases at a finite rate, $\lambda = \omega_2 t$,
such that the system is  subject to {\it quasiperiodic} driving by two  modes with  incommensurate frequencies $\omega_1$ and $\omega_2$.
Defining $\tilde{\mathcal{H}}(\phi_1,\phi_2)\equiv \hat{\mathcal{H}}_0(\phi_1/\omega_1; \phi_2)$, the Hamiltonian can hence be written as 
$\hat{\mathcal{H}}(t) = \tilde{\mathcal{H}}(\omega_1 t,\omega_2 t )$, 
where $\tilde{\mathcal H}(\phi_1,\phi_2)$ is $2\pi$-periodic in each of its arguments.
Note that the discussion below applies  to any quasiperiodically-driven 1d system of noninteracting fermions whose Hamiltonian  can be expressed in this form. 

Due to the absence  of interactions, time-evolved many-body states   in the system can be resolved in terms of 
 Slater determinants  of time-evolved single-particle states. 
For simplicity,  below  we therefore consider the dynamics of the system with only a single particle present, unless otherwise stated.
We use calligraphic symbols to denote many-body operators (acting in Fock space), and italic symbols for single-particle operators.

{\it Frequency localization ---}
{As a main result, this work shows that the model above is characterized by an integer-valued topological invariant when it is localized in the spatial and frequency domains}. 
The key to understanding such localization  is   a generalized Floquet theorem~\cite{Ho_1983,Quasi_Luck88,Quasi_Casati89,Quasi_Blekher92,Hatami_2016,Else_2020}:
for {the} bichromatically 
 driven system{s}{ we consider},  a complete orthonormal basis of generalized  (single-particle) Floquet states $\{|\Phi_n(\phi_1,\phi_2)\rangle\}$ can be defined such that the time-evolution  of any state takes the form
$
|\psi(t)\rangle = \sum_n  \kappa _n e^{-i\varepsilon_n t}|\Phi_n(\omega_1 t , \omega_2 t )\rangle .
$
Here each $|\Phi_n(\phi_1,\phi_2)\rangle$ is $2\pi$-periodic in each of its arguments while   $\varepsilon _n$ is real-valued and defines a generalized quasienergy.
The structure above 
is equivalently captured in 
  the single-particle evolution operator of the system, $\hat U(t) \equiv \mathcal T e^{-i\int_0^t \!{\rm d}t'  \hat{H}(t')}$, where $\mathcal T$ denotes {the} time-ordering {operation} and $\hat {H}(t)$ is the single-particle Hamiltonian of the system (i.e., $\hat{\mathcal{H}}(t)$ restricted to the one-particle sector).
Specifically,   
\be
 \hat U (t) = \hat {P}(\omega_1 t,\omega_2 t) e^{-i\hat{H}_{\rm eff}t},
\label{eq:micromotion}
\ee
where $\hat{P}(\phi_1,\phi_2) \equiv \sum_n |\Phi_n(\phi_1 ,\phi_2 )\rangle\langle \Phi_n(0,0)|$ and $ \hat{H}_{\rm eff} \equiv \sum_n \varepsilon _n |\Phi_n(0,0)\rangle\langle \Phi_n(0,0)|$ 
define a generalized micromotion operator and effective Hamiltonian for the system, respectively.

The decomposition in 
Eq.~(\ref{eq:micromotion}) 
is only useful if each generalized Floquet state $|\Phi_n(\phi_1,\phi_2)\rangle$ is 
a continuous  function of  $\phi_1$ and $\phi_2$, or, equivalently, if the two-dimensional Fourier decomposition of $|\Phi_n(\phi_1,\phi_2)\rangle$  converges.
This situation defines 
 ``frequency localization.'' 
With disorder, the generalized Floquet states may moreover be {\it spatially} localized~\cite{SOM}, implying that  particles remain confined near their initial location at all times.
We refer to the combination of 
 spatial and frequency localization as  ``full localization'' below. 

To infer the conditions for frequency localization, we work in the Fourier harmonic space corresponding to mode $2$, yielding the Hamiltonian  of an effective {\it two-dimensional}, {\it periodically driven} system, $\hat{\boldsymbol H}_{\rm 2d}(t) = \hat{\boldsymbol H}_{\rm 2d}(t+T_1)$.
To this end we introduce a new degree of freedom, $\hat n_2$, whose corresponding ``Fourier harmonic''  Hilbert space is spanned by the states $\{|y\rangle_{\rm F},y\in \mathbb{Z}
\}$,  such that $\hat n_2|y\rangle_{\rm F} = y|y\rangle_{\rm{F}}$.
Heuristically, $\hat n_2$  can be seen as counting the number of photons in mode $2$~\cite{Shirley1965_1}.
We obtain $\hat{\boldsymbol H}_{\rm 2d}(t)$ from $\hat H(t)$ by adding a  term $\omega_2 \hat n_2$ and replacing each phase factor $e^{i\omega_2 t}$ in $\hat H(t)$ by $\sum_{y=-\infty}^\infty |y\rangle_{\rm F} \langle y-1|_{\rm F}$ {(and similar for the corresponding Hermitian conjugate, $e^{-i\omega_2 t}$)}.
See Supplementary Online Material (SOM) for further details.

The Hamiltonian $\hat{\boldsymbol H}_{\rm 2d}(t)$  acts on the Hilbert space  spanned by the states $|x,y\rrangle \equiv |x\rangle\otimes |y\rangle_{\rm F}$ where $y\in \mathbb{Z}$,  $x=1,\ldots 2L$, and $|x\rangle$ denotes the single-particle state of  the original 1d system with the particle  located on site $x$. 
Thus, $\hat{\boldsymbol H}_{\rm 2d}(t)$ can be seen as the 
 Hamiltonian of a  two-dimensional lattice system whose sites are indexed by $x$ and $y$.
Each Floquet state of $\hat {\boldsymbol H}_{\rm 2d}(t)$, $|\Psi_n(t)\rrangle$,   corresponds to a generalized Floquet state of $\hat H(t)$, $|\Phi_n(\phi_1,\phi_2)\rangle$, via $\langle x|\Phi_n(\phi_1,\phi_2)\rangle = \sum_{y}  e^{-iy\phi_2}\llangle x,y|\Psi_n(\phi_1/\omega_1)\rrangle $~\cite{Ho_1983,SOM}. 
The $y$-dependence of $|\Psi_n(t)\rrangle$ thus encodes the Fourier components of $|\Phi_n(\phi_1,\phi_2)\rangle$ with respect to $\phi_2$. 
Hence, full localization corresponds to localization of  the Floquet states of $\hat {\boldsymbol H}_{\rm 2d}(t)$  due to  Anderson localization in the spatial direction and Wannier-Stark localization in the frequency (Fourier harmonic) 
 direction~\cite{SOM}.
 
The above considerations imply that frequency localization requires  
   irrational $\omega_2/\omega_1$: 
when $\omega_2/\omega_1$ is sufficiently close to $p/q$ for some integers $p$ and $q$, the oscillating terms of ${\boldsymbol H}_{\rm 2d} (t)$ resonantly couple sites separated by $q$ lattice constants in the $y$-direction, inducing $y$-delocalization, and hence frequency delocalization after translation back to the   Hilbert space of the physical (1d) problem at hand ~\cite{Quasi_Blekher92,Else_2020,SOM}.
We thus expect frequency localization to   break down in some  $\omega_2$-interval around $p\omega_1/q$  for each choice of  integers $p$ and $q$.
However, the width of this interval may decrease with increasing $p$ and $q$, allowing  frequency localization to occur for a finite-measure set of $\omega_2$~\cite{SOM}.

{\it Topological invariant ---} 
The topological invariant of the QFTEP can be defined from the generalized micromotion operator  in Eq.~(\ref{eq:micromotion}). 
For simplicity,  we consider a system with periodic boundary conditions; the results can be applied directly to systems  with open boundary conditions.
\fnrm{As a first step, we}
We first define a ``phase-twisted'' micromotion operator $\hat{P}_{\theta}(\phi_1,\phi_2)$ by adding a factor $e^{- i \theta}$ ($e^{i\theta}$) to the matrix elements of $\hat P(\phi_1,\phi_2)$ that transfer a particle across an arbitrary reference bond $x_0$ in the positive (negative) $x$-direction~\cite{SOM}.
When  the system is  fully localized,  
 $\hat{P}_\theta(\phi_1, \phi_2)$ is unitary, as well as continuous and periodic in  
 $\theta$, $\phi_1$, and $\phi_2$~\cite{SOM}. 
Under these conditions, $\hat{P}_\theta(\phi_1, \phi_2)$ is characterized by an integer valued winding number: 
\be 
\nu
 = \!
  \oint\!\frac{{\rm d} \theta {\rm d}^2 \phi}{8\pi^2}  
  \Tr\Big(\!\hat{ P}_\theta^{-1} \partial _\theta\hat{ P}_\theta \, [  \hat{ P}_\theta^{-1} \partial _{\phi_2}\hat{ P}_\theta\, , \, \hat{ P}_\theta^{-1} \partial _{\phi_1}\hat{ P}_\theta ]\Big),
\label{eq:WnDef}
\ee
where ${\rm d}^2\phi = {\rm d}\phi_1 {\rm d}\phi_2$, and  we suppressed the phase dependence of $\hat P_\theta(\phi_1,\phi_2)$ for brevity.
The index $\nu$ cannot change under smooth deformations of the system parameters that preserve 
full localization, and thus defines the invariant of the QFTEP. 
Nonzero values of $\nu$ can arise for weak or moderate disorder, where particles  undergo  nontrivial  micromotion while their dynamics remain localized on long length scales.


The invariant $\nu$ can be seen as a dimensional reduction
 of the winding number of the anomalous Floquet-Anderson insulator (AFAI)~\cite{FQI2,FQI3}.
Recall that full localization of $\hat H(t)$ occurs when the Floquet eigenstates of $\boldsymbol{\hat H}_{\rm 2d}(t)$ are  localized. 
In this case, $\boldsymbol{\hat H}_{\rm 2d}(t)$ is characterized by  the integer-valued winding number of the AFAI~\cite{FQI2,FQI3}. 
A  straightforward derivation shows that this winding number 
 is identical to $\nu$~\cite{SOM}.

{\it Bulk-edge correspondence ---}
For a system with open boundary conditions, a nonzero value of $\nu$ implies a quantized transport of  energy  between modes $1$ and $2$ when all sites near one edge are occupied.
Using the correspondence between Floquet states of $\Hes_{\rm 2d}(t)$ and generalized Floquet states of $\Hat H(t)$, in the SOM we show that the time-averaged rate of  work done on mode $2$, 
$\hat {\mathcal P}_2(t)\equiv  -\omega_2 \partial_{\phi_2} \tilde {\mathcal H}(\omega_1 t, \omega_2 t)$,
is quantized when the system initially has 
sites $1,\ldots x_1$ occupied, for some $x_1$ in the bulk of the chain:
\be 
\lim_{\tau \to \infty} \frac{1}{\tau}\int_0^\tau\!\!{\rm d}t\, \langle \hat {\mathcal P}_2(t)\rangle= \nu \mathcal{P}_0.
\label{eq:bec}
\ee
Conservation of energy dictates that the average rate of work done on mode $1$ is given by $-\nu\mathcal P_0$. 
Since Eq.~\eqref{eq:bec} is independent of $x_1$, only fermions near the edge contribute to the quantized energy transfer above.
This result 
establishes the bulk-edge correspondence of the QFTEP. 

To understand Eq.~\eqref{eq:bec}, note that for a finite open chain, $\Hes_{\rm 2d}(t)$ describes an AFAI on an infinite strip in the $y$-direction {(corresponding to ``photon number'' of mode 2)}. 
The bulk-edge correspondence for the AFAI dictates that $\Hes_{\rm 2d}(t)$ supports chiral modes propagating along  the edges of the strip that carry a quantized average current $\nu/T_1$ and $-\nu/T_1$ along the $y$-direction at the left and right end of the strip, respectively~\cite{FQI2}. 
According to the mapping through which $y$ corresponds to  the number of photons absorbed by mode $2$,     we   
    expect that particles on the left edge transfer energy to mode $2$  at the quantized  average rate $\nu \mathcal{P}_0$.

The  quantized energy transfer to mode $2$ is supported by topologically protected edge modes: 
through the   relationship  between the QFTEP and AFAI established above, 
the existence of chiral edge states in the  AFAI 
 implies that each end of the  QFTEP supports a family of generalized Floquet states which are  localized spatially but delocalized in frequency space.
These families of states, or ``edge modes,'' are topologically protected features that can only disappear if full localization is destroyed in the bulk. 
 When all  states in one such topological edge  mode are occupied, they collectively generate a quantized flow of energy to mode $2$ at the  rate $\pm\nu \mathcal P_0$.
Away from full occupation, we expect quantization  persists if the particle density, $p(x)$, is locally uniform over the characteristic localization length scale of generalized Floquet states~\cite{Ye_2020}. 
In this case, the  left and right topological edge modes are occupied with probability $p(0)$ and $p(L)$, respectively, resulting in the energy transfer rate $\nu \mathcal P_0[p(0)-p(L)]$.


\begin{figure}
\includegraphics[width=0.99\columnwidth]{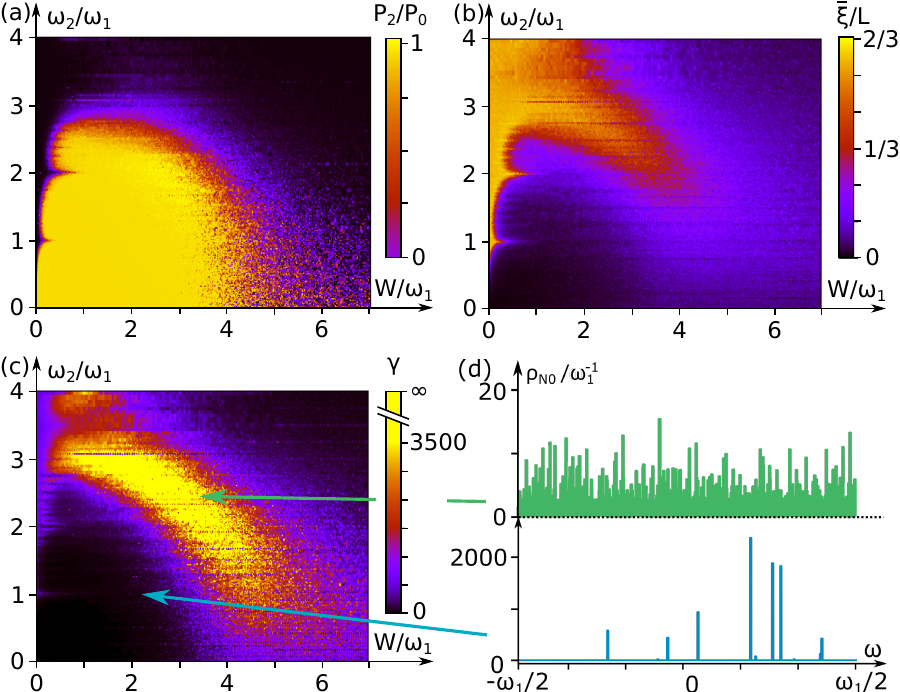}
\caption{Simulation of the QFTEP  with  $200$ unit cells and open boundary conditions, as detailed in the numerics section. 
(a) Average energy absorption rate over $10\hspace{1.5pt}000$ periods of mode $1$ when the left half of the chain is initially occupied. 
(b)  Estimated   {spatial} localization length of the generalized Floquet states, $ \xi$. 
(c)  Effective spectral density of the system, $\gamma $. 
(d) Spectral density $\rho_{N_0}(\omega)$ resulting from a wave-packet initialized on a single site in the middle of the chain for parameters within the topological plateau ({bottom}), and at the transition ({top}).}
\label{f2}
\end{figure}
\emph{Numerical simulations ---} 
We  demonstrate the quantized energy pumping of  the QFTEP through numerical simulations of the model presented in the introduction. 
We simulated a system of  $L=200$ unit cells (
$400$ sites) with open boundary conditions, initialized by only filling the leftmost 200 sites with fermions.
Using direct  time evolution, we computed the time-averaged rate of energy transfer to  mode $2$, $\bar{\mathcal P}_2$, by averaging $\langle \hat {\mathcal P}_2(t)\rangle$ over the  time-interval from $0$ to $t_{\rm f}  = 10,\!000\, T_1$. 
Fig.~\ref{f2}(a) shows $\bar {\mathcal P}_2/\mathcal{P}_0$, as a function of $\omega_2$ and $W$~\cite{omega_sampling}. 
The data {indicate} 
 a large plateau of  quantized energy pumping ($\bar{\mathcal P}_2 = \mathcal{P}_0$) at finite values of $\omega_2$ and $W$, supporting our conclusions above.

The data in Fig.~\ref{f2}(a) exhibit several  features that are consistent with our discussion above. 
The plateau with $\bar{\mathcal P}_2=\mathcal P_0$ has maximal extent along the $\omega_2$-direction when $W\approx \omega_1$, 
indicating that weak to moderate disorder stabilizes the QFTEP (interestingly, for  $\omega_2 \ll \omega_1 $,  stabilization occurs for very weak disorder).
The  quantized plateau diminishes when $\omega_2/\omega_1$ approaches values $ p/q$ for  integers $p$ and $q$, as is particularly clear for $q=1$, where the yellow plateau region is sharply ``pinched in.''
In the SOM we provide high-resolution data confirming that quantization of $\bar{\mathcal P}_2$  breaks down for other values of $p$ and $q$, with the  breakdown most pronounced   for smaller  $p$ and $q$.
These data thus support our prediction that the QFTEP is protected by the 
irrationality of  $\omega_2/\omega_1$.

{To characterize the phase transitions of the QFTEP, we}  
estimate the spatial localization length  of the generalized Floquet states in the system via
$ \xi   \equiv \sup_{x_1} \ ( \sum_{x_2}   |\langle x_1|\hat U(t_{\rm f})|x_2\rangle|^2(x_1-x_2)^2)^{1/2}$.
Fig.~\ref{f2}(b) shows $\xi$ for the same parameters as taken in Fig.~\ref{f2}(a). 
For $W \lesssim 4\omega_1$, the boundary of the topological plateau in Fig.~\ref{f2}(a) clearly coincides with 
the region in Fig.~\ref{f2}(b) where $\xi\sim \mathcal O (L)$ (indicating delocalization).
For $\omega_2,W\gtrsim 3\omega_1$ we observe a localized,  topologically-trivial phase with $\xi \sim \mathcal{O}(1)$ and $\bar{\mathcal{P}}_2 = 0$ [Figs.~\ref{f2}(a\mkadd{-}b)]. 
Rather than a direct transition to this topologically trivial phase, Fig.~\ref{f2}(b) indicates the existence of an intermediate delocalized region for $W\lesssim 3\omega_1$, $\omega_2\gtrsim 3\omega_1$. 


When $W$ exceeds $3\,\omega_1$, the topological phase transition changes its qualitative behavior: 
$\xi$ decreases and $\bar {\mathcal P}_2$ displays irregular fluctuations for a  finite  $\omega_2$-interval.
To {investigate} 
the transition here, we computed a measure of the {\it frequency-space} localization length, 
using  the time-evolution of a single-particle state initialized on a particular  site $x_{\rm i}$ in the chain, $|\psi(t)\rangle \equiv \hat U(t)\Ket{x_{\rm i}}$.
For a given (large) integer $N$, we let $|\psi_N(\omega)\rangle \equiv \frac{1}{\sqrt{\omega_1 N}}\sum_{n=1}^N e^{in\omega T_1} |\psi(n{T_1})\rangle$ denote the finite-time discrete Fourier transform of $|\psi(t)\rangle$ when sampled stroboscopically with 
the period of mode 1.
We then define an effective spectral density $\gamma$ characterizing $\Ket{\psi(t)}$ via the
inverse participation ratio of the normalized spectral distribution  $\rho_N(\omega) \equiv |\langle \psi_N(\omega)|\psi_N(\omega)\rangle|$: $ \gamma \equiv\frac{1}{\omega_1}\lim_{N\to \infty} N / [\int_0^{\omega_1}\!\!{\rm d\omega} \rho_N(\omega)^2]$.
When  $|\psi(t)\rangle$ has a  dense 
Fourier  spectrum, $\int_0^{\omega_1}\!\!{\rm d\omega} \rho_N(\omega)^2$ remains finite in the limit $N\to \infty$, and thus $\gamma $ diverges.
However, for a discrete spectrum, where
    $\rho_\infty(\omega) \equiv \lim_{N\to \infty} \rho_N(\omega)$ is given by a discrete sum of delta functions, the integral diverges linearly with $N$, such that $\gamma $ remains finite.
In this case, 
$\gamma$ gives the inverse sum of the squared peak weights in $\rho_\infty(\omega)$; i.e., it measures effective number of peaks in the frequency spectrum  of $|\psi(t)\rangle$ (modulo $\omega_1$), $\rho_\infty(\omega)$. 
We hence expect that  $\gamma$ is  a good proxy for  the localization length in the frequency space of mode $2$.

Fig.~\ref{f2}(c) shows the maximal value  of $\gamma $ 
 obtained from 
 $15$ time-evolved single-particle states with initial positions $x_{\rm i}$ randomly chosen within
the middle third of the  system. We used the same parameters as considered in panels (a-b), 
and set $N$ to $N_0=10\zs000$.
As an illustration, in Fig.~\ref{f2}(d), we plot  $\rho_{N_0}(\omega)$ for two parameter sets [indicated with arrows in Fig.~\ref{f2}(c)] where $\gamma $ is large (upper) and small (lower), respectively.
Although Fig.~\ref{f2}(b) does not conclusively indicate whether spatial delocalization is present   for $W\gtrsim 3\omega_1$, 
Fig.~\ref{f2}(c) shows that the system undergoes frequency delocalization throughout the entire topological phase boundary. 


{\it Realization in  a driven two-level system --- }
Here we propose {a dimensionally-reduced} experimental
realization of the  QFTEP in a two-level system (qubit) driven by three incommensurate modes. 
The  
 model  presented in the introduction 
is mapped to this platform 
by taking the limit of zero disorder, and replacing  spatial crystal momentum   with the phase of a third driving mode: $e^{ik} \to e^{i\omega_3 t}$~\cite{Qi_2008}.
When the frequencies $\omega_{1},\omega_2$, and $\omega_3$ are sufficiently incommensurate,  the generalized Floquet eigenstates of the system remain localized due to the quasi-disorder of the lattice in the three-dimensional frequency space~\cite{Ho_1983,Quasi_Luck88,Quasi_Casati89,Quasi_Blekher92,Hatami_2016,Else_2020,SOM}.
The  analysis below Eq.~\eqref{eq:micromotion} 
thereby also yields the topological index $\nu$ for this model, defined by Eq.~\eqref{eq:WnDef} with $\theta$ replaced by the phase $\phi_3 = \omega_3 t$ of the third driving field.
In the SOM we  provide data from numerical simulations that  confirm the 3-mode driven qubit model described above supports a  topologically nontrivial regime  characterized by  $\nu=1$.
Edges can naturally be incorporated into the qubit realization \fnrm{above}
  by replacing one or more of the driving modes  by  quantized cavity modes, whose vacuum states define a natural edge.

{\it Summary and outlook ---} 
This work establishes  the QFTEP as a new non-equilibrium topological phase. 
The QFTEP elevates the  Floquet-Thouless energy pump (FTEP) in Ref.~\cite{TQEP1} from a fine-tuned (but nonetheless interesting) phenomenon to a  genuine stable 
 phase of matter.
Whereas the FTEP requires adiabatic driving, fine-tuning, and is formally destroyed by disorder~\cite{TQEP1}, the QFTEP  arises for finite $\omega_1$ and $\omega_2$, and is {\it stabilized} by disorder. 
In a setting with open boundary conditions, the QFTEP is characterized by robust, quantized energy pumping between the two driving modes, supported by topological edge states. \fnrm{ that appear at the ends of the chain. }

We \fnrm{further} showed that a qubit driven by three incommensurate modes can realize a dimensionally-reduced version of the QFTEP. 
\fnrm{In the short term, s}
Such a  system is a promising platform for experimental realization of this phenomenon due to well-developed techniques for controlling and driving qubits~\cite{OC}.
The QFTEP  may also be directly realized in one-dimensional quantum chains, such as systems of ultracold atoms in optical lattices, or trapped-ion systems.
The robustness to disorder and finite modulation frequency  makes such  realizations of the QFTEP more feasible than for the adiabatic FTEP. 

Quantization of energy pumping breaks down for rational values of $\omega_1/\omega_2$,
implying that the phase diagram of the QFTEP has a fractal structure. 
Understanding this fractal structure and its role for the phase transitions of the QFTEP  will  be an interesting direction for future studies. 


Another prospect for invesigations is to apply  our dimensional reduction scheme 
 to other equilibrium or Floquet topological phases of matter. 
 A more complete investigation of  the physical signatures of the QFTEP, its stability to interactions, and its place in the expanding classification of non-equilibrium topological phases of matter will also be important future directions.
   
%
{\it Acknowledgements ---}
We thank Anushya Chandran, Philip Crowley, David Long, Ivar Martin, and Gil Refael for valuable discussions. 
This work was performed with support from the National Science Foundation through award number DMR-1945529 and the Welch Foundation through award number AT-2036-20200401 (M.K. and R.G.).  S.G. acknowledges support from the Israel Science Foundation, Grant No. 1686/18.
F.N.~and M.R.~gratefully acknowledge the support of the European Research Council (ERC)
under the European Union Horizon 2020 Research and
Innovation Programme (Grant Agreement No. 678862), and the Villum Foundation.
We used the computational resources of the Lonestar 5 cluster operated by the Texas Advanced Computing
Center at the University of Texas at Austin and the Ganymede and Topo clusters operated by the University of Texas at Dallas' Cyberinfrastructure \& Research Services Department.

{\it Note added by the authors ---}
During the preparation of this manuscript,  a preprint appeared which describes an  energy pumping phenomenon related to the one we consider here~\cite{Long_2020}, in the context of a  general classification. 
 Our work is   fully consistent with  Ref.~\cite{Long_2020}, and  provides a complementary perspective on the phenomenon, including a study of the
role of spatial disorder, along with additional experimental proposals for realizing the phase.
\bibliography{energy_pump_bibliography}

\begin{thebibliography}{49}
\bibitem [{ome()}]{omega_sampling}%
  \BibitemOpen
  \href@noop {} {}\bibinfo {note} {Each value of $\omega_2$ we probed was
  obtained by adding a small random number to a value generated by a different
  algorithm. Consequently, each sampled value of $\omega_2$ was incommensurate
  with $\omega_1$ (up to machine precision) for all but a measure-zero set of
  outcomes}\BibitemShut {NoStop}%
\bibitem [{bou()}]{boundary_conditions}%
  \BibitemOpen
  \href@noop {} {}\bibinfo {note} {A natural concern is that it is unnatural to
  introduce periodic boundary conditions along the $\hat n_2$ ($y$) direction
  due to the constant tilt $\omega_2 \hat n_2$. However, when the Floquet
  states are localized in the $\hat n_2$-direction, we can arbitrarily modify
  the behavior at $\hat n_2 \to \pm \infty$ without affecting the topological
  properties for finite $\hat n_2$.}\BibitemShut {Stop}%
\end{thebibliography}
\onecolumngrid
\vspace{0.8cm}
\begin{center}
\textbf{\large Supplementary Material}
\vspace{0.5cm}
\end{center}
\maketitle
\twocolumngrid

\section{Extended Hilbert space and frequency localization}
\label{app:ehs}
In this section, we provide details of the extended Hilbert space approach that was used in the main text (See also, e.g.,  Refs.~\cite{Ho_1983,Quasi_Luck88,Quasi_Casati89,Quasi_Blekher92,Hatami_2016,Else_2020}).
In Sec.~\ref{seca:ehs} we define the extended Hilbert space Hamiltonian that results from including the Fourier harmonic (or ``frequency'') space of mode $2$, $\Hes_{\rm 2d}(t)$, and  show how the generalized Floquet states of $\hat H(t)$ can be obtained from the Floquet states of $\Hes_{\rm 2d}(t)$. 
We subsequently demonstrate how frequency localization 
 is equivalent to spatial localization of the Floquet states of $\Hes_{\rm 2d}(t)$ (Sec.~\ref{seca:floc}).

\subsection{Extended Hilbert space}
\label{seca:ehs}
As explained in the main text,  the generalized Floquet states of  $\hat H(t)$,  $\{\Ket{\Phi_n(\phi_1,\phi_2)}\}$,
 can be obtained from the Floquet states of  a {\it periodically driven}, {\it two-dimensional} system with Hamiltonian $\Hes_{\rm 2d}(t)$~\cite{Ho_1983,Quasi_Luck88,Quasi_Casati89,Quasi_Blekher92,Hatami_2016,Else_2020}. 
The  extra dimension in this ``extended space'' corresponds to the \fnrm{integer-valued}Fourier harmonic index ($q$) in  the Fourier decomposition of  $\Ket{\Phi_n(\omega_1 t, \omega_2 t)}$ with respect to mode $2$: 
\be 
\Ket{\Phi_n(\phi_1,\phi_2)} = \sum_{q} e^{iq\phi_2}\Ket{\Phi_n^{(q)}(\phi_1)}.
\label{eqa:floquet_decomp}
\ee

To construct $\Hes_{2{\rm d}}(t)$ we introduce an effective position operator $\hat{n}_2$ in the Fourier harmonic space of mode $2$;  as explained in the main text, this operator can  heuristically be seen as  counting the number of photons in mode $2$~\cite{Shirley1965_1}. The energy in mode $2$ transforms to an effective linear potential along the corresponding lattice direction, $\omega_2\hat n_2$, and each factor of $e^{-i m\omega_2 t}$ in $\hat{H}(t)$ induces an $m$-th neighbor hopping term in the corresponding direction, 
$e^{-i\omega_2 mt} \to \hat{B}^{(m)}_2 = \sum_{y=-\infty}^\infty |y\rangle \langle y+m|$, where $\hat n_2 |y\rangle = y |y\rangle$. 
Applying this transformation thus yields:  
\be 
\Hes_{2{\rm d}}(t) = \omega_2 \hat n_2+ \sum_{q} \hat H^{(q)}(t)  \hat B_2^{(q)},
\label{eqa:ehs_ham}
\ee
where $\hat H^{(q)}(t)$ denotes the $q$-th Fourier component of the single-particle Hamiltonian with respect to the phase of mode $2$,  $\hat H^{(q)}(t) \equiv \frac{1}{2\pi}\int_{0}^{2\pi}\!\!{\rm d}\phi_2\, \tilde H(\omega_1 t,\phi_2) e^{iq\phi_2}$, where $\tilde H(\phi_1,\phi_2)$ denotes the single-particle Hamiltonian of the  system as a function of the two modes' phases (i.e., the restriction of $\tilde {\mathcal H}(\phi_1,\phi_2)$  the single-particle sector). 
In the following we use double brackets, $|\cdot\rrangle$, for states that live in the extended Hilbert space on which $\Hes_{\rm 2d}(t)$ acts.

The Hamiltonian $\Hes_{\rm 2d}(t)$ describes a two-dimensional single-particle system, where the $x$-direction corresponds to the original (spatial) coordinate, and the $y$-direction to the Fourier harmonic index $n_2$. 
Evidently, $\Hes_{\rm 2d}(t)$ has periodic time-dependence with the period of mode $1$, $T_1$: $\Hes_{\rm 2d}(t) = \Hes_{\rm 2d}(t+T_1)$.
As a result, the time-evolution generated by $\Hes_{\rm 2d}(t)$ can be decomposed in terms of a complete orthonormal basis of Floquet states $\{|\Psi_n(t)\rrangle\}$ that are time-periodic with period $T_1$: $|\Psi_n(t)\rrangle  = |\Psi_n(t+T_1)\rrangle$~\cite{Shirley1965_1}.
%
For any Floquet state of $\Hes_{\rm 2d}(t)$, $|\Psi_{n}(t)\rrangle$,  it is possible to identify a unique generalized Floquet state of $\hat H(t)$, $|\Phi_{n'}(\phi_1,\phi_2)\rangle$ such that, for any state $|\alpha \rangle$ in the 1d system:
\be 
\langle \alpha |\Phi^{(q)}_{n'}(\omega_1 t)\rangle = \llangle \alpha ,q|\Psi_{n}(t)\rrangle,
\label{eqa:fes_correspondence}
\ee
where $|\alpha,n_2 \rrangle \equiv |\alpha \rangle \otimes |n_2\rangle$ denotes the  basis state in the extended Hilbert space  at location $y=n_2$ corresponding to $|\alpha \rangle $~\cite{Ho_1983,Quasi_Luck88,Quasi_Casati89,Quasi_Blekher92,Hatami_2016,Else_2020}. 
This establishes the relationship between the Floquet states of $\Hes_{\rm 2d}(t)$ and the generalized Floquet states of $\hat H(t)$.

Due to the built-in symmetry of $\Hes_{\rm 2d}(t)$, $\hat B^{(1)}_2 \Hes_{\rm 2d}(t) [\hat B^{(1)}_2]^\dagger = \Hes_{\rm 2d} +\omega_2$,  the Floquet states of $\Hes_{\rm 2d}(t)$ can be grouped into families whose elements are mapped to each other through shifts in $n_2$:
 if $|\Psi_n(t)\rrangle$ is a Floquet state of $\Hes_{\rm 2d}(t)$ with associated quasienergy $\varepsilon _n$, for each $m$,  $\hat B_2^{(m)}|\Psi_n(t)\rrangle$ also is a Floquet state of $\Hes_{\rm 2d}(t)$, with quasienergy $\varepsilon _n - m\omega_2$. 
Through Eq.~\eqref{eqa:fes_correspondence} each such family of  Floquet states of $\Hes_{\rm 2d}(t)$ correspond   to the same generalized Floquet state of $\hat H(t)$, 
 up to a $\phi_1$-dependent phase factor. 

The above mapping can be also applied to mode $1$. 
In particular, the operation that maps $\hat H(t)$ to  $\Hes_{\rm 2d}(t)$ in Eq.~\eqref{eqa:ehs_ham} can subsequently be applied to the latter, resulting a {\it time-independent} Hamiltonian of a {\it 3-dimensional} system, where the third dimension corresponds to the Fourier harmonic index with respect to mode  $1$~\cite{Shirley1965_1}. 
The same approach can moreover  be 
 generalized to systems driven quasiperiodically by $3$ or more modes~\cite{Ho_1983,Quasi_Luck88,Quasi_Casati89,Quasi_Blekher92,Hatami_2016,Else_2020}. 

\subsection{Frequency  localization}
\label{seca:floc}
We now show that frequency localization  of $\hat H(t)$  is equivalent to spatial localization of the Floquet states of $\Hes_{\rm 2d}(t)$. 

As described in the main text, frequency localization arises when each generalized Floquet state of $\hat H(t)$, $|\Phi_n(\phi_1,\phi_2)\rangle$, is continuous in its arguments $\phi_1$ and $\phi_2$. 
Such continuity occurs when  $|\Phi_n(\phi_1,\phi_2)\rangle$ can be approximated arbitrarily well by  truncating its Fourier decomposition, 
\be 
|\Phi_n(\phi_1,\phi_2)\rangle = \sum_{p,q} e^{ip \phi_1 +i q\phi_2 }|\Phi_n^{(p,q)}\rangle,
\label{eqa:multi_fourier}
\ee
 at some sufficiently large, but finite, cutoff in the Fourier Harmonic indices $p$ and $q$.
Note that the boundedness of $\hat H(t)$ [and hence also $\partial _t U(t)$], along with  Eq.~(1) 
of the main text, implies that, as a function of $p$ and $q$,  $|\Phi_{n}^{(p,q)}\rangle$ may can only be significant  near  a particular line where $p\omega_1 + q\omega_2$ takes some (given) constant value. 
Consequently, 
%
frequency localization is ensured
when $|\Phi_n^{(p,q)}\rangle$ effectively only has significant support for a finite range of $q$.
To  be more precise, frequency localization occurs when the squared norm of  $|\Phi_n^{(p,q)}\rangle$ 
for large $q$ falls off at least as fast as $|q|^{-\alpha}$ for some $\alpha >1$. (The condition can equivalently be expressed   in terms of   $p$.) 
In this sense, frequency localization with respect to mode $2$ implies frequency localization with respect to mode $1$, and vice versa.

To relate frequency localization to the localization properties of $\Hes_{\rm 2d}(t)$, we note from Eq.~\eqref{eqa:fes_correspondence} that 
\be 
\langle \alpha |\Phi^{(p,q)}_{n'}\rangle = \frac{1}{T_1}\int_0^{T_1}\!\!{\rm dt} e^{-i\omega_1 t}\llangle \alpha ,q|\Psi_{n}(t)\rrangle.
\ee
From the discussion above, we conclude that frequency localization arises when the Floquet states of $\Hes_{\rm 2d}(t)$, $\{|\Psi_n(t)\rrangle\}$, are spatially localized in the $y$-direction (corresponding to the Fourier Harmonic index of mode $2$, $q$).

We expect frequency localization automatically implies {\it full} localization (i.e., localization of the generalized Floquet states of $\hat H(t)$ both in real and frequency space):
for a frequency localized system, we expect the  Floquet states of $\Hes_{\rm 2d}(t)$ can be obtained to arbitrarily high precision by truncating the  extended Hilbert space at some sufficiently large cutoff in $n_2$.
Since the   resulting system is effectively one-dimensional,  the Floquet states of $\Hes_{\rm 2d}(t)$  are subject to Anderson localization~\cite{Shirley1965_1}, and hence will also be spatially localized.

{
\section{Role of incommensurability}
}\begin{figure}
\includegraphics[width=0.99\columnwidth]{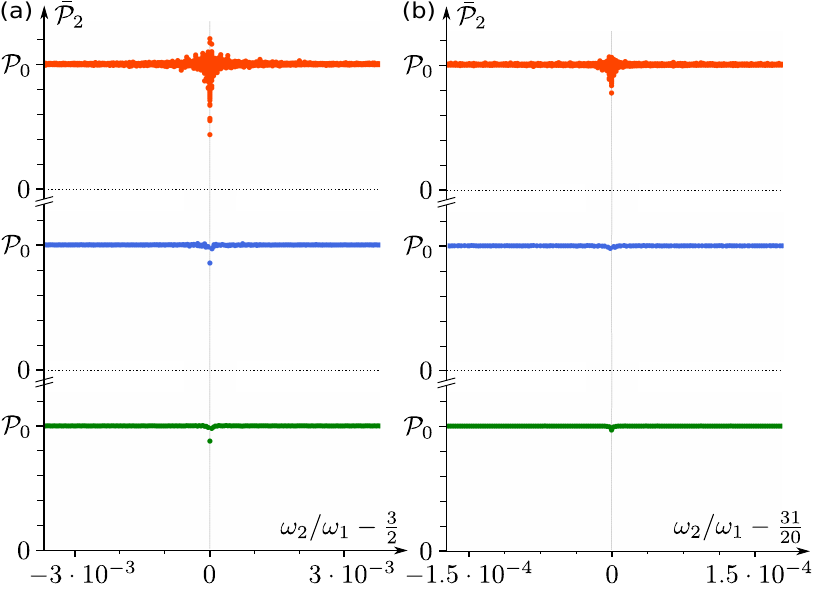}
\caption{
Breakdown of quantization of energy pumping near rational frequency ratios. 
Left and right panels show the average rate of energy transfer to mode $2$, $\bar{\mathcal P}_2$, in a QFTEP model with disorder strength $W=2\omega_1$,  computed for 
 different values of  $\omega_2$ in the vicinity of $\frac{3}{2} \omega_1$ (a) and $ \frac{31}{20} \omega_1$ (b).
For each probed value of $\omega_2$,  $\bar{\mathcal P}_2$ is computed for several  disorder realizations and  values of the initial phase difference between the driving modes, $\Delta \phi$. 
 Red data points (top) show the  values of $\bar{\mathcal P}_2$ obtained from individual disorder realizations with randomly chosen  $\Delta \phi$, 
blue data points (middle) show the  value of $\bar{\mathcal P}_2$ after averaging over $\sim 10$ disorder realizations  with  $\Delta \phi=0$, and green data points (bottom) show the  value of $\bar{\mathcal P}_2$ after averaging  over $\sim 80$ disorder realizations with randomly chosen $\Delta \phi$. 
  See main text for further details of the model and simulation. 
Note the different scales on the $x$-axes  in panels (a) and (b).
}
\label{fig:commensurate_plot}
\end{figure}

Here we provide additional details of the  relationship between frequency localization and  the  incommensurability of $\omega_1$ and $\omega_2$.
We moreover provide data from numerical simulations to demonstrate that the quantization of the energy pumping rate, $\bar{\mathcal P}_2$, breaks down near rational values of $\omega_2/\omega_1$.

The condition of irrationality of  $\omega_2/\omega_1$ can be understood from the extended Hilbert space picture: 
when $\omega_2/\omega_1$ is sufficiently close to $p/q$ for some integers $p$ and $q$, the oscillating terms of ${\boldsymbol H}_{\rm 2d} (t)$ resonantly couple sites separated by $q$ lattice constants in the $y$-direction, inducing $y$-delocalization, and hence frequency delocalization~\cite{Quasi_Blekher92,Else_2020}.
 For each pair of integer values of $p$ and $q$, frequency localization  hence breaks down  in some finite $\omega_2$-interval around $\omega_2 = p\omega_1/q$, $I_{pq}$.

We expect that the width of  $I_{pq}$, $|I_{pq}|$, decays as a function of $p$ and $q$, with a rate controlled by the harmonic structure of $\tilde H(\phi_1,\phi_2)$. 
Specifically, if the $(p,q)$th Fourier coefficient of $\tilde H(\phi_1,\phi_2)$, $\tilde H_{pq}$, is nonzero,  sites in the 2d lattice separated by  $(p,q)$  lattice constants are coupled directly by $\Hes_{\rm 2d}(t)$, and we expect $|I_{pq}|$ to scale with $|\tilde H_{pq}|$. 
 If on the other hand the Fourier  coefficient $\tilde H_{mn}$ is only significant for a few small values of $m$ and $n$ (i.e., if the $\tilde H(\phi_1,\phi_2)$ is a smooth function of its arguments),  the resonant coupling for large $p,q$ is induced by high-order virtual processes.
 In this case we  expect  $|I_{pq}|$ to decrease exponentially as a function of $p$ and $q$.

The decay of $|I_{pq}|$ with $p$ and $q$ means that frequency localization can occur for a finite-measure set of $\omega_2$ even though it breaks down for a dense set of points.
Additionally, it is more likely to occur for highly irrational values of $\omega_2/\omega_1$, i.e., when $\omega_2/\omega_1$  is not close to  $p/q$ for some small integers $p$ and $q$~\footnote{See, e.g., Ref.~\cite{Leinonen_2017} for a review of irrationality measures.}.

To support the discussion above, we simulated the QFTEP model  introduced   in the main text  [see Fig.~1(c)] for values of $\omega_2$ near rational multiples of $\omega_1$.
We chose the disorder strength $W=2\omega_1$, which places the system within the $\nu=1$ QFTEP phase    for $0<\omega_2\lesssim 2.5\omega_1$ [see Fig.~2(a) in the main text].
For  values of $\omega_2$ which we specify below,  we computed the average rate of energy transfer to mode $2$, $\bar{\mathcal P}_2$, over a time-interval of length $10\, 000\,T_1$, after initializing the system with the leftmost $200$  sites occupied by fermions and the remaining sites  empty (i.e., we performed the same calculation as used for Fig.~2 in the main text).

For each value of $\omega_2$ we probed, we computed $\bar{\mathcal P}_2$ for multiple disorder realizations, and for multiple   values of the initial phase difference between the two modes, $\Delta \phi$.
Here $\Delta \phi$ controls the offset of the second argument of  $\tilde{\mathcal H}$, such that the Hamiltonian of the system in our simulation is given by $\hat{\mathcal H}(t) = \tilde{\mathcal H}(\omega_1t,\omega_2t + \Delta \phi)$ (see main text for definition of $\hat{\mathcal H}$ and $\tilde{\mathcal H}$).
For incommensurate frequencies, different values of $\Delta \phi$ correspond to shifts of time-origin, and thus lead to the same value of $\bar{\mathcal P}_2$ after  averaging over sufficiently long time intervals. 
For commensurate or near-commensurate frequencies (i.e., frequencies that are effectively commensurate over the time-window we probe), $\bar{\mathcal P}_2$ generally depends on  $\Delta \phi$, which is why we introduce this parameter in our simulations.

We  first computed  $\bar{\mathcal P}_2$  for $\omega_2/\omega_1$ near the rational value $3/2$.
The corresponding data are plotted in Fig.~\ref{fig:commensurate_plot}(a). 
For each value of $\omega_2$ probed in Fig.~\ref{fig:commensurate_plot}(a), we computed $\bar{\mathcal P}_2$  for $\sim80$ different disorder realizations and randomly selected values of $\Delta \phi$, with each value shown as a separate data point in red. 
For most probed $\omega_2$-values,  $\bar{\mathcal P}_2$ takes the quantized value $\mathcal P_0$ for all disorder realizations. 
However, when $\omega_2/\omega_1$ approaches $3/2$, the quantization   clearly breaks down. 
We estimate the $\omega_2$-interval where the quantization of $\bar{\mathcal P}_2$  breaks down for individual realizations  to have width of order $5*10^{-4}\omega_1$.
The breakdown of quantization persists after averaging  $\bar{\mathcal P}_2$ over individual disorder realizations with the value of  $\Delta \phi$  fixed to zero (blue data points), or averaging over disorder realizations {and} values of $\Delta \phi$ (green). 
The data thus agree well with our expectation  that the quantization of $\bar{\mathcal P}_2$ breaks down near rational values of $\omega_2/\omega_1$.

Next, we  computed  $\bar{\mathcal P}_2$  for $\omega_2/\omega_1$ near the ``less rational'' value $31/20$, using the same method as above. 
The corresponding data are plotted in Fig.~\ref{fig:commensurate_plot}(b) [note that the $x$ axis is rescaled by a factor $20$ relative to  the $x$-axis in Fig.~\ref{fig:commensurate_plot}(a)]. 
As Fig.~\ref{fig:commensurate_plot}(b) shows, $\bar{\mathcal P}_2$ clearly deviates from the quantized value $\mathcal P_0$ when $\omega_2/\omega_1$ approaches  $31/20$.
However, the deviation from quantization is not as pronounced as in panel (a), and  the breakdown  occurs in a much smaller interval of $\omega_2$.
We estimate the the $\omega_2$-interval  where quantization breaks down for individual realizations to have width of  order $10^{-5}\omega_1$, and thus be a factor $\sim  50$ smaller than the corresponding interval surrounding $\omega_2= \frac{3}{2}\omega_1$.
These features thus agree  with our expectation that the breakdown of quantization of $\bar{\mathcal P}_2$  near rational frequency ratios $\omega_2/\omega_1=p/q$  is more pronounced at highly rational frequency ratios (i.e., for small $p$ and $q$) than at less rational frequency ratios (i.e., for large $p$ and $q$).

\section{Unitarity of $\hat P_\theta (\phi_1,\phi_2)$}
\label{seca:phase_twist_unitarity}Here we show that ``phase-twisted'' micromotion operator $\hat P_\theta (\phi_1,\phi_2)$ [introduced above Eq.~(2) 
in the main text] 
 is unitary when the generalized Floquet states of the system are spatially localized on the scale of the system size, $L$.
We used this result in the main text to establish the integer quantization of the invariant $\nu$ that characterizes the QFTEP. 

We recall that  $\hat P_\theta (\phi_1,\phi_2)$ is constructed from the generalized micromotion operator in Eq.~(1) 
 by appending the
phase factor $e^{ -i \theta}$ ($e^{i\theta}$) to matrix elements of $\hat P(\phi_1,\phi_2)$ that hop a particle across a cut at $x=x_0+1/2$ in the positive (negative) $x$-direction, where $x_0$ is some arbitrary reference site.
In the case of periodic boundary conditions, the hopping trajectory between two sites  is defined to be the shortest path between the sites on the lattice.
Thus, letting $|\alpha \rangle$ denote the single-particle state where the particle is located on site $\alpha $, $\hat P_\theta(\phi_1,\phi_2)$ can be written: 
\be 
\hat P_\theta (\phi_1,\phi_2) = \sum_{\alpha ,\beta } |\alpha \rangle \langle \alpha |\hat P (\phi_1,\phi_2)|\beta \rangle\langle \beta | e^{-is_{\alpha \beta } \theta}, 
\ee
where  $s_{\alpha \beta } $ takes value $1$ ($-1$) if the shortest path from site $\alpha $ to site $\beta $ traverses the cut at $x=x_0+1/2$ in the positive (negative) $x$-direction, and takes value $0$ if the path does not traverse this cut.
In particular, $s_{\alpha \alpha }=0$.
In the case of periodic boundary conditions (when the  generalized Floquet states are localized on the scale  $L$), the  operation above is equivalent to inserting a flux $\theta$ through the closed chain formed by the system,  with a  gauge where the corresponding phases are applied {\it locally}~\cite{FQI2}.

To establish the unitarity of  $\hat P_\theta$ 
we first note that  
\be
\langle \alpha_1 |\hat P_\theta^\dagger \hat P_\theta |\alpha _2\rangle = \sum_\beta\langle \alpha_1 |\hat P^\dagger|\beta \rangle\langle \beta |\hat P |\alpha _2\rangle e^{-i\theta(s_{\alpha _1\beta } -s_{\alpha _2 \beta })},
\label{eqa:unitary_1}
\ee
where, here and below, we suppress the dependence of $\hat P_\theta (\phi_1,\phi_2)$ and $\hat P(\phi_1,\phi_2)$ on $\phi_1,\phi_2$  for brevity.

From the definition of $\hat P$  below Eq.~(1) 
 in the main text, we see that  $\langle \alpha |\hat P|\beta \rangle$ is only nonzero when sites $\alpha $ and $\beta $ are located within a distance of order $\xi $ from each other, where $\xi$ denotes the spatial localization length of the generalized Floquet states.
Using our assumption that $\xi$ is much smaller than the system size, $L$, one  can confirm that $s_{\alpha _1\beta } -s_{\alpha _2 \beta } = s_{\alpha _1 \alpha _2}$ for all contributing 
 terms in the sum in Eq.~\eqref{eqa:unitary_1}. 
Using this in Eq.~\eqref{eqa:unitary_1}, we obtain
\be
 \langle \alpha_1 |\hat P_\theta^\dagger \hat P_\theta |\alpha _2\rangle =\langle \alpha_1 |\hat P^\dagger\hat P |\alpha _2\rangle e^{-i\theta s_{\alpha _1\alpha _2 }}.
\ee
Since $\hat P$ is unitary,  $\langle \alpha _1|P^\dagger P|\alpha _2\rangle = \delta_{\alpha _1\alpha _2}$, where $\delta_{ab}$ denotes the Kronecker symbol. 
Using that $s_{\alpha \alpha }=0$, we find  $\langle \alpha _1|\hat P_\theta^{\dagger} \hat P_\theta^\dagger|\alpha _2\rangle =\delta_{\alpha _1\alpha _2}$. 
Thus we  conclude that $\hat P_\theta$ is unitary, which was the goal of this section.

\section{Relationship with AFAI}
\newcommand{\Pt}{\bs{\hat P}}
Here we show that  the q-FTEP invariant $\nu$
is identical to the AFAI invariant of the extended Hilbert space Hamiltonian of the system,  $\Hes_{\rm 2d}(t)$, which was introduced below Eq.~(1) 
in the main text~\cite{FQI2}. 
To be more specific, recall from the main text  $\nu$ is robustly quantized  when the generalized Floquet states of $\hat H(t)$
 are localized both in position and frequency space. 
Such full localization 
is equivalent to spatial localization of the Floquet states of $\boldsymbol{\hat H}_{\rm 2d}(t)$. 
When such localization occurs, $\boldsymbol{\hat H}_{\rm 2d}(t)$ is characterized by  the integer-valued winding number of the AFAI, $\nu_{\rm AFAI}$,~\cite{FQI2,boundary_conditions}. 
The goal of this section is to show that $\nu = \nu_{\rm AFAI}$. 

To  establish that $\nu=\nu_{\rm AFAI}$, we note that the latter 
 can be defined from   an analogous expression to Eq.~(2) from the main text 
  using the $T_1$-periodic micromotion operator of $\Hes_{\rm 2d}(t)$, 
$
 \Pt(t) \equiv \sum_n |\Psi_n(t)\rrangle \llangle \Psi_n(0)|,
 $  where $\{|\Psi_n(t)\rrangle\}$ denote the Floquet states of $\Hes_{\rm 2d}(t)$  (see Sec.~\ref{app:ehs}).
Analogously to the ``phase-twisted'' micromotion operator $\hat P_\theta(\phi_1,\phi_2)$  from the main text, we construct the phase-twisted micromotion operator $\Pt_{\theta_1,\theta_2}(t)$ by  appending the phase factor  $e^{ -i \theta_x}$  ($e^{i\theta_x}$) to matrix elements of $\Pt(t)$ that hop a particle across a cut at $x=x_0+1/2$ in the positive (negative) $x$-direction, and similarly assign $e^{\mp i \theta_y}$  to the matrix elements that hop across a cut at $y=y_0 + 1/2$, for some arbitrary reference sites $x_0$ and $y_0$. 
From $\Pt_{\theta_1,\theta_2}(t)$, we can compute $\nu_{\rm AFAI}$  as follows~\cite{TQEP2}:
\be 
\nu_{\rm AFAI}  
 = \!
  \oint\!\frac{{\rm d}^2 \theta {\rm d}t}{8\pi^2}  
  \Tr\Big(\!\bs{\hat{ P}}^{-1} \partial _{t}{\bs{\hat{ P}}} \, [  \Pt^{-1} \partial _{\theta_1}\bs{\hat{ P}}\, , \, \bs{\hat{ P}}^{-1} \partial _{\theta_2}\bs{\hat{ P}}]\Big),
\ee
where we suppressed the dependence of $\Pt_{\theta_1,\theta_2}(t)$ on $\theta_1,\theta_2$, and $t$ in the above for brevity.
%
%
%
%
Through the correspondence between the generalized Floquet states of $\hat{H}(t)$ and the Floquet states of ${\Hes}_{\rm 2d}(t)$ in Eq.~\eqref{eqa:fes_correspondence}, it can be shown by direct insertion that $\nu_{\rm AFAI}$ is identical to $\nu$, as defined in Eq.~(2) of the main text. 

\section{Bulk-edge correspondence}
Here we establish  Eq.~(3) 
 in the main tex, which gives the  bulk-edge correspondence  for the quasiperiodic Floquet-Thouless energy pump:
we show that in a setting with open boundary conditions, when the system is  initialized  by filling all sites from the left end  up to a given (arbitrary) site located well within the bulk, $x_{1}$, the average rate of energy transfer from the system to mode $2$ is given by $\nu \mathcal P_0$, where $\mathcal P_0 = 2\pi\omega_1\omega_2$ denotes the ``quantum of energy transfer.''
Specifically, we show that 
\be 
\lim_{\tau \to \infty} \frac{1}{\tau}\int_0^\tau\!\!{\rm d}t\, \langle \hat {\mathcal P}_2(t)\rangle= \nu \mathcal{P}_0,
\label{eqa:bec_app}
\ee
where, as in the main text, the many-body operator $\hat {\mathcal P}_2(t)$ denotes the rate of energy transfer to  mode $2$ from the system: $\hat{\mathcal P}_2(t) \equiv  -\omega_2 \partial_{\phi_2} \tilde {\mathcal H}(\omega_1 t, \omega_2 t)$, where $\tilde {\mathcal H}(\phi_1,\phi_2)$ denotes the many-body Hamiltonian of the system as a function of the phases of the two modes (see also main text).

To obtain Eq.~\eqref{eqa:bec_app}, we recall  that time-evolved many-body states in the system are given by Slater determinants of time-evolved single-particle states (due to the absence of interactions). 
Hence   the evolution  resulting from the initialization  above can  be decomposed in terms of the evolved  single-particle states  $\{|\psi_x(t)\rangle\}$, $x=1\ldots x_1$, where $|\psi_x(t)\rangle$  describes the evolution resulting from initializing the system with a single  fermion on site $x$.  
In particular,
\be
\langle \hat {\mathcal P}_2(t)\rangle =
\sum_{x=1}^{x_1} \omega_2  F_{2,x}(t),
\ee
where $F_{2, x} =  -\langle \psi_x(t)| \partial _{\phi_2} \tilde { H}(\omega_1t,\omega_2t)|\psi_x(t)\rangle$
denotes the generalized force  exerted on mode $2$ by the system in the state $|\psi_x(t)\rangle$, and $\tilde H(\phi_1,\phi_2)$ denotes the single-particle Hamiltonian as a function of the two modes' phases (i.e., the restriction of $\tilde{\mathcal H}(\phi_1,\phi_2)$ to the single-particle sector). 

Through the correspondence between the generalized Floquet eigenstates of $\hat H(t)$ and the Floquet states of $\Hes_{\rm 2d}(t)$ in Eq.~\eqref{eqa:fes_correspondence}, 
one can verify that the long-time-averaged value of $F_{2,x}(t)$, $\bar F_{2,x} \equiv  \lim_{\tau \to \infty}\frac{1}{\tau} \int_0^\tau\!{\rm d}t\, F_{2,x}(t)$, gives the time-averaged $y$-velocity, $\bar v_y(x,0)$ of a particle initially localized on the site $(x,0)$ in the two-dimensional system described by $\Hes_{\rm 2d}(t)$.
From the built-in symmetry of $\Hes_{\rm 2d}(t)$ (see Sec.~\ref{seca:ehs}) one can moreover verify that $\sum_{x=1}^{x_1} \bar v_y(x,0)$ gives the time-average of the total $y$-current, $\bar I_y$, passing a given cut along the $x$-axis in the two-dimensional fermionic system governed by the many-body generalization of  $\Hes_{\rm 2d}(t)$, when all sites  located at $x\leq x_0$ are occupied.
The bulk-edge correspondence of the AFAI dictates that $\bar I_y=\nu/T_1$~\cite{FQI2,FQI3}.
Combining the above results, we conclude that the average work done on mode $2$ by the system, when initialized by filling sites $1\ldots x_1$, is given by Eq.~\eqref{eqa:bec_app}.
This was what we set out to show.

\section{Numerical simulation of trichromatically driven qubit}

\begin{figure}
\includegraphics[width=0.99\columnwidth]{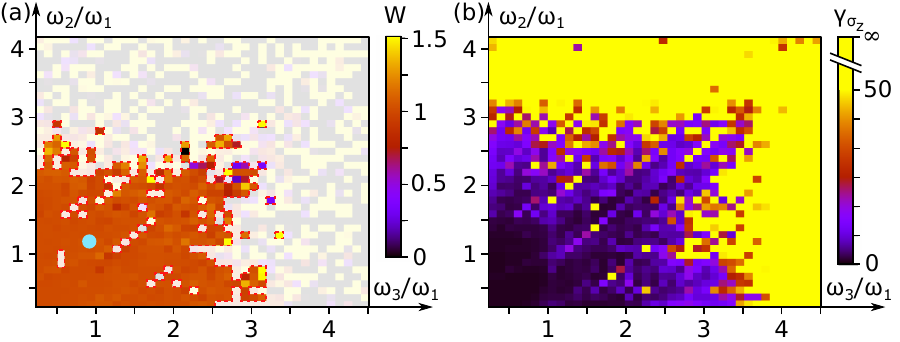}
\caption{A trichromatically-driven qubit realization of the q-FTEP.
(a) Topological index, $\nu$, obtained directly from Eq.~(2) in the main text, as a function of $\omega_2$ and $\omega_3$.
The white-shaded region indicates parameters where the effective spectral density in panel (b) exceeds $10$ and the simulation  may not faithfully capture the system's dynamics. 
(b) Effective spectral density for  the model, $\gamma_{\sigma_z}$. }
\label{f3}
\end{figure}
Here we  provide data from numerical simulations of the 3-mode driven qubit model described in the main text. 
The simulations confirm that this system supports a  topologically nontrivial regime  characterized by the $3$-mode generalization of the invariant $\nu$ taking value $1$  (see main text for definition).

To demonstrate the nontrivial topology of  the trichromatically-driven qubit, 
we simulate a  qubit Hamiltonian 
similar to  the dimensionally-reduced version of $\hat H_{\rm dr}(t)$ 
from the introduction. The 
 Hamiltonian  we consider  features an additional fifth step where it takes value $\omega_1 \sigma^z$, where $\sigma^{x,y,z}$ denote Pauli operators acting on the qubit states.
 We  obtain the generalized Floquet eigenstates for the system by extending the Hilbert space to include the frequency space of modes $2$ and $3$~\cite{Ho_1983,Quasi_Luck88,Quasi_Casati89,Quasi_Blekher92,Hatami_2016,Else_2020}.
To make this tractable, we truncated the  Fourier harmonic indices 
 of these  modes   to  $N_{\rm c}=50$ each.

To investigate the extent of frequency localization in the system, we computed 
$\tilde\sigma_{z;N}(\omega) = \sum_{n=1}^N \frac{e^{in\omega T_1}}{\omega_1 N}\langle \frac{\sigma_z(nT)+1}{2}\rangle$ with $N=65\zs 536$, and initial  state ${|\!\uparrow\rangle}$.
We then obtained the effective spectral density  $\gamma _{\sigma_z} =  (\int_0^{\omega_1}\!\!{\rm d}\omega \,|\tilde\sigma_{z;N}(\omega)|)^2/ \int_0^{\omega_1}\!\!{\rm d\omega}\, |\tilde \sigma_{z;N}(\omega)|^2$.
We expect $\gamma _{\sigma_{z}} $ serves as an equivalently valid measure of frequency localization  to $\gamma $,  
while being more  accessible for  experimental measurement.
For $\omega_2,\omega_3 \lesssim 3\omega_1$, we found $\gamma _{\sigma_z}<N_c$, indicating frequency localization. 

We computed the topological invariant $\nu$ from the $3$-mode generalization of Eq.~(2) from the main text, using the generalized micromotion operator for the system [see Fig.~\ref{f3}(a)].
The white color in Fig.~\ref{f3}(a) indicates the region where $\gamma _{\sigma_z}$  exceeds $10$, in which the simulation is not expected to accurately capture the system's dynamics. 
In the localized region, $\nu$ exhibits a well-defined plateau where it takes value $1$, indicating that the system is in a topologically nontrivial regime.

\end{document}